**Dissecting the properties of neutron star–black hole mergers originating in dense star clusters**

*Manuel Arca-Sedda*[1]

[1]Astronomisches Rechen-Institut, Zentrum für Astronomie der Universität Heidelberg, 69120 Heidelberg, Germany (m.arcasedda@gmail.com)

**The detection of gravitational waves emitted during a neutron star - black hole merger and the associated electromagnetic counterpart will provide a wealth of information about stellar evolution nuclear matter, and General Relativity. While the theoretical framework about neutron star - black hole binaries formed in isolation is well established, the picture is loosely constrained for those forming via dynamical interactions. Here, we use *N*-body simulations to show that mergers forming in globular and nuclear clusters could display distinctive marks compared to isolated mergers, namely larger masses, heavier black holes, and the tendency to have no associated electromagnetic counterpart. These features could represent a useful tool to interpreting forthcoming observations. In the Local Universe, gravitational waves emitted from dynamical mergers could be unravelled by detectors sensitive in the decihertz frequency band, while those occurring at the distance range of Andromeda and the Virgo Cluster could be accessible to lower-frequency detectors like LISA.**

The third observational campaign, O3, operated by the LIGO-Virgo Collaboration (LVC) will enlarge the family of gravitational wave (GW) sources, currently comprised of 11 confirmed black hole (BH) binary and 1 neutron star (NS) binary mergers[1]. The loot of observations accumulated during O3 will hopefully include the first BH binary with component masses in the so-called upper mass-gap, i.e. $50 - 150$ M$_\odot$, and the first NS-BH merger. In fact, as reported in the gravitational-wave candidate event database (GraceDB, https://gracedb.ligo.org/), both events might have been already recorded. The observation of an NS-BH merger represents a crucial cornerstone in both GW astronomy and stellar evolution and dynamics. The GW signal emitted by this type of sources encodes information about the mass ratio of the binary, the BH spin, the NS compactness and equation of state[2–4]. The next generation of GW detectors will allow us to chase these objects up to the dawn of Universe[5], offering us the unique chance to follow them from the inspiraling phase to the merger. The formation of an accretion disc during the merger could trigger a kilonova event[6] and power a short Gamma-ray burst (sGRB)[7], making NS-BH mergers promising multi-messenger sources. The coincident observation of GWs and a kilonova[8], the detection of peculiar precession in the jets produced during the sGRB[9], or the anisotropic emission of ejected matter[10] are some of the proposed signatures of a putative NS-BH electromagnetic (EM) counterpart. The properties of the EM emission depend on the binary properties close to the merger. A high eccentricity, for instance, can affect the amount of mass ejected, the mass accreted onto the BH and the angular momentum transferred[11]. The actual development of an EM counterpart is expected to depend likely on the mass ratio, the BH spin, and the NS equation of state[2–4,10–12]. For mass ratios smaller than $\simeq 1/3 - 1/4$, the NS undergoes tidal disruption inside the BH's innermost stable circular orbit (ISCO), thus preventing the EM emission[13].

The scenarios behind NS-BH formation are still not fully understood. In the case of isolated binaries, binary population synthesis tools predict typical merger rates $\Gamma = 9 - 100$ yr$^{-1}$ Gpc$^{-3}$ at redshift zero[14,15], nearly circular mergers[16] with typical chirp masses[17] $\mathcal{M}_{\rm chirp} = 3$ M$_\odot$, and BHs with masses[18] strongly peaked around $m_{\rm BH} \sim 7$ M$_\odot$ and, in general, smaller than 20 M$_\odot$. The picture is loosely constrained for NS-BH mergers formed via dynamical interactions in star clusters (GCs), nuclear clusters (NCs) and galactic nuclei[19], owing to the recent advances in our understanding of the physics governing the formation and retention of BHs and NSs. Indeed, star clusters might be able to retain long-lived BH subsystems[20,21,22] -- which could persist at present in a number of Milky Way GCs[22,23], the detection of NS in star clusters suggest natal kicks lower than previously thought[24,25], and the discovery of GWs emitted by BHs as heavy as 30 $M_\odot$ revolutionized our knowledge of stellar evolution for single and binary stars. Large number of BHs and NSs and the presence of heavy BHs can impact significantly the probability for NS-BH binary to form and, possibly, merge. Bridging reliable stellar dynamics, up-to-date stellar evolution recipes, and a detailed description of the last phases of binary evolution is crucial to assess the properties of dynamical mergers. Finding significant differences between dynamical and isolated mergers would represent a piece of crucial information to interpret future GW observations.

Here, we study the complex dynamical interactions involving BHs and NSs in dense clusters, focusing on hyperbolic encounters between a binary, composed of a compact object and a stellar companion, and a single compact object, exploring two configurations: either the compact object in the binary is a NS and the third object is a BH (configuration NSSTBH), or vice-versa (BHSTNS). Combining our simulations with observations of Galactic GCs and NC in the local Universe, and with Monte Carlo simulations of GCs, we infer for dynamical NS-BH mergers an optimistic merger rate of 0.1 events per year and Gpc cube in the case of GCs, and 0.01 events per year and Gpc cube for NCs. Despite the small value, we find that dynamical mergers exhibit peculiarities that make them distinguishable from isolated mergers: chirp masses above 4 $M_\odot$, BH masses above 20 $M_\odot$, and the absence of associated electromagnetic emission if the BH is highly spinning and has a mass above 10 $M_\odot$. We calculate the associated GW emission showing that these mergers can be observed with detectors sensitive in the decihertz band and even with millihertz detectors like the laser interferometer space antenna (LISA), provided that they took place at distances typical of the Andromeda Galaxy or the Virgo galaxy cluster.

**Results**

**Dynamical formation of NS-BH binaries in star clusters.** To investigate this dynamical formation channel we exploit 240000 direct *N*-body simulations that take into account up-to-date stellar evolution recipes for natal BH mass[26] and General Relativistic corrections in the form of post-Newtonian formalism[27]. Figure 1 shows the trajectories of one of our simulations. As detailed in the Method section, we vary both scattering parameters (binary semimajor axis and eccentricity, impact parameter and velocity of the third object) and environmental quantities (metallicity and velocity dispersion of the host cluster). To connect our results to real star clusters, we exploit the catalogue of Milky Way GCs to find a tight relation connecting the GC velocity dispersion ($\sigma$), mass ($M_{GC}$), and half-mass radius ($R_{GC}$) (see the Method section for more details). To complement and support our simulations, we perform a deep analysis on the MOCCA Survey Database I[28], a collection of over 2000 Monte Carlo models of globular clusters that span a wide portion of the phase space and represent globular clusters with present-day masses $M_{GC} \sim 3\times10^5 M_\odot$ and half-mass radii $R_{GC} \sim 1\text{-}3$ pc . This sample allowed us to reconstruct the history of all NSSTBH and BHSTNS in 1298 models, and to derive an average scattering rate of $dR_{sca}/dt = 6.3$ Gyr$^{-1}$ for configuration NSSTBH and 245.4 Gyr$^{-1}$ for BHSTNS. We find that a scattering results in the formation of a NS-BH in ~1.27 - 1.59% of the cases, with the lower(upper) value corresponding to the NSSTBH(BHSTNS) configuration, but none of them merge within a Hubble time. On average, for configuration BHSTNS the scatterings occur at ~0.01 times the cluster core radius $R_c$, whereas for NSSTBH the scattering location is broadly distributed between 0.01 and 0.3 $R_c$, but still well inside the cluster interiors. These scatterings occur late in the cluster life, usually several times the cluster half-mass relaxation time $t_{rel}$. Figure 2 shows the distribution of the scattering time, $t_{sca}$, normalized to $t_{rel}$ , calculated at 12 Gyr.

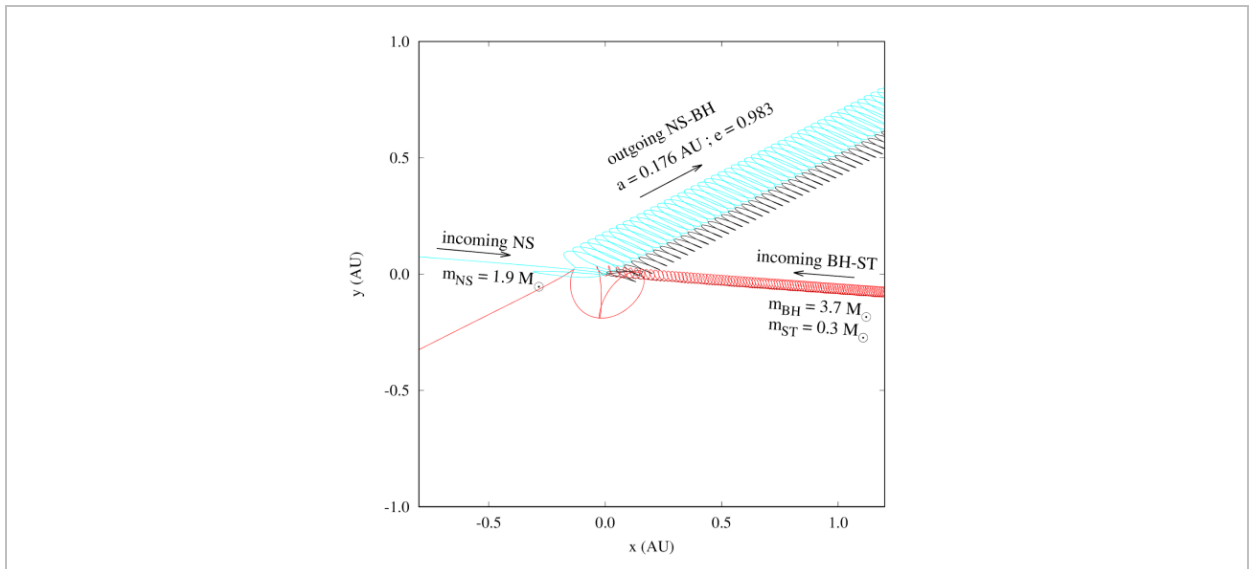

Fig 1. Interaction between a black hole – star binary (BH-ST) and a single neutron star (NS). This sketch shows the trajectories in the x-y plane (in astronomical units, AU) of a NS (cyan line) with a mass of $m_{NS} = 1.9 M_\odot$ scattering over a BH-ST binary (red and black lines) with masses $m_{BH} = 3.7\ M_\odot$ and $m_{ST} = 0.3\ M_\odot$. After the interaction, a NS-BH binary forms with a final semimajor axis $a = 0.176$ AU and eccentricity $e = 0.983$.

**The probability of NS-BH mergers.** To identify potential mergers in our 240000 *N*-body models we need to associate to any newborn binary a formation time, $t_{form}$. This is calculated through two quantities: the scattering time $t_{sca}$, which we extract from the distribution of $t_{sca}/t_{rel}$ derived for MOCCA models (see Figure 2), and the cluster relaxation time $t_{rCL}$, which we extract from the distribution of values calculated for 157 Galactic GCs[29] and 228 NCs[30]. The NS-BH formation time is thus calculated as $t_{form} = t_{sca} / t_{rel}$ x $t_{rCL}$. Despite the richness of information encoded in the MOCCA database, current models represent GCs and cannot be used to describe more extreme environments like NCs. Moreover, stellar evolution for BHs is not updated yet and the treatment used for close encounters does not include General Relativistic corrections. Therefore, we use *N*-body simulations to perform a thorough investigation of this dynamical channel, using the analysis performed on MOCCA to: a) compare with the scattering rate derived by *N*-body simulations, and b) infer the time at which a scattering can occur. After formation, we calculate the NS-BH merger time[31] $t_{GW}$ and, for each candidate, we draw 100 different values of $t_{form}$, retaining only candidates for which the drawings is $t_{form} + t_{GW} < 14$ Gyr in at least 50% of the cases. Unfortunately, this requirement alone does not ensure that a merger can successfully take place. Indeed further interactions can soften and even destroy it if the NS-BH binding energy is lower than the mean kinetic energy of the environment[32]. The limiting value of the binary semimajor axis above which this can happen is called hard-binary separation $a_h = G(m_1+m_2)/\sigma^2$.

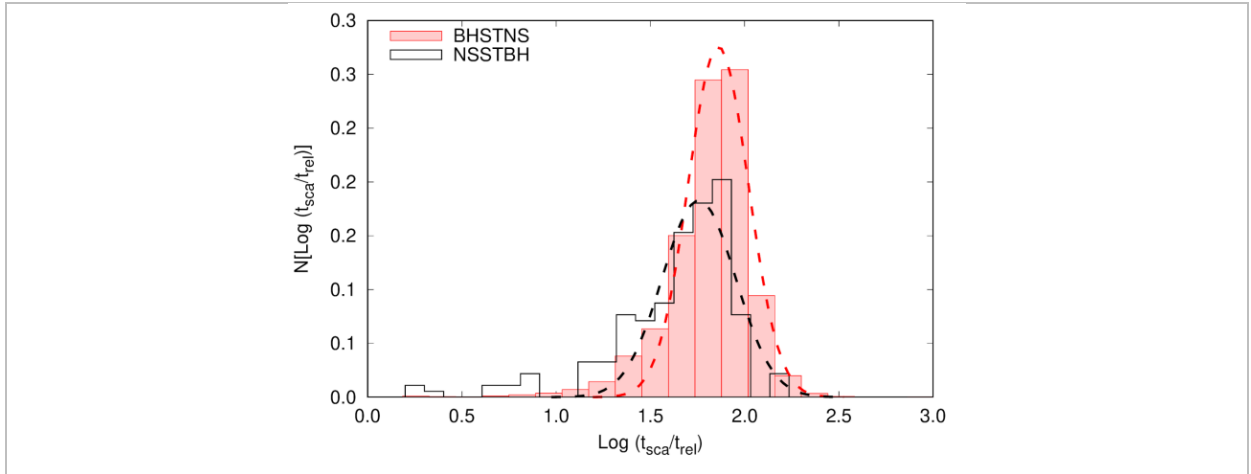

Fig 2. Distribution, in logarithmic values, of the ratio between the scattering time ($t_{sca}$) and the relaxation time ($t_{rel}$). The distribution refers to the scatterings observed in MOCCA Monte Carlo models for the configuration representing a black hole-star binary interacting with a neutron star (configuration BHSTNS, red filled boxes) or viceversa (configuration NSSTBH, black open steps). Dotted lines identify the best-fit Gaussian curve. The distribution is well described by a Gaussian with best-fit values for the mean $m = 0.725 \pm 0.008$ and dispersion $s = 0.174 \pm 0.008$ (shown as a dotted red curve for BHSTNS), and $m = 0.66 \pm 0.02$ and $s = 0.15 \pm 0.02$ (shown as a dotted black curve for NSSTBH).

"Soft" binaries, $a > a_h$, can be disrupted by strong encounters over an evaporation time[34] $t_{evap} = [\sigma^3(m_1+m_2) / (32\pi G^{1/2} 2 m_* \rho\, a \ln \Lambda)]^{1/3}$, depending on the binary properties ($m_1, m_2, a$), the cluster velocity dispersion ($\sigma$), density ($\rho$), and average stellar mass ($m_*$), and the Coloumb logarigthm ($\ln \Lambda$). Therefore, to shortlist merger candidates we require the simultaneous fulfillment of three conditions: i) $t_{form} + t_{GW} < 14$ Gyr, to avoid NS-BH binaries with delay times larger than a Hubble time, ii) $a > a_h$, to avoid soft NS-BH binaries, iii) $t_{GW} < t_{evap}$, to avoid NS-BH binaries that can be disrupted by further interactions.

Note that the delay time calculated this way don't account for the cluster formation time, $t_{fCL}$, thus among all candidates satisfying simultaneously the three conditions above only a fraction $f$ will satisfy also $t_{fCL} + t_{form} + t_{GW} < 14$ Gyr. In our calculations, we assume that the majority of clusters form at redshift $z \sim 2$[34], corresponding to $t_{fCL} \sim 10$ Gyr. As shown in Figure 3a, the fraction of merging NS-BH, $p_{GW}$, increases at increasing the sigma, but depends poorly on the scattering configuration (NSSTBH or BHSTNS) and the metallicity. A rough limit to the merger rate in the local Universe for NS-BH mergers in clusters can be written as[35]:

$$\Gamma = f \Gamma_c \rho_{MWEG} N_c, \qquad (1)$$

where $\Gamma_c$ is the merger rate per unit of time and cluster, $\rho_{MWEG} = 0.0116$ Mpc$^{-3}$ is the local density of galaxies[36], and $N_c$ is the number of clusters in a given galaxy. The merger rate per cluster is given by

$$\Gamma_c = N_{bin}\, p_{GW}\, dR/dt \qquad (2)$$

where $dR/dt$ is the rate of binary-single interactions and can be calculated combining $N$-body and MOCCA models as detailed in the Method section. To infer the number of binaries that at a given time co-exist in the cluster we exploit the 12 Gyr output of MOCCA models, in the case of NSSTBH configuration a GC hosts up to 4 NS-stellar binaries in 90% of the cases, and up to 7 binaries in the remaining 10%, while for BHSTNS GCs have < 4 BH-stellar binaries in the 95% of the cases, and up to 12 in the remaining 5%. In our calculations we assume $N_{bin} = 4$ as a fiducial value. As shown in Figure 3b, $\Gamma_c$ increases at increasing the velocity dispersion, is larger for the BHSTNS configuration at fixed sigma value, and larger for lower metallicities. In all the cases, $\Gamma_c$ is well described by a power-law in the form $\Gamma_c = (\sigma / \sigma_c)^\alpha$. Configuration BHSTNS displays a larger $\Gamma_c$ values due to the fact that they involve heavier binaries compared to NSSTBH, thus they are characterized by larger cross section and, thus, scattering rates.

Using the $\Gamma_c$-$\sigma$ dependence we can exploit Equation 1 to calculate the merger rate for Milky Way equivalent galaxies, namely those galaxies that share similar properties with our own, like a population of $N_c \sim 200$ metal poor clusters with a relatively low velocity dispersion, $\sigma \sim 5$-$6$ km/s.

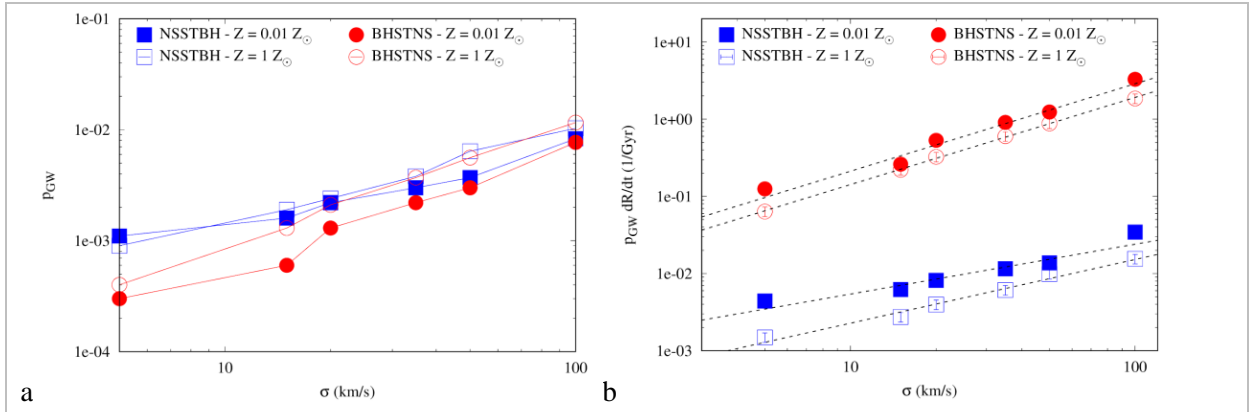

Fig 3. Neutron star – Black hole (NS-BH) mergers formation probability. **a.** Probability for NS-BH mergers for different configurations. The plot shows the fraction of models producing a NS-BH merger as a function of the cluster velocity dispersion $\sigma$ in kilometers per second. Interactions between a black hole – star binary and a single neutron star (BHSTNS) are marked with circles, whereas the viceversa (NSSTBH) are marked with squares. Open symbols refer to a metallicity value $Z = 0.01\ Z_\odot$ (being $Z_\odot$ the typical solar value), whereas filled symbols refer to $Z = 0.01\ Z_\odot$. **b.** Merger rate per single cluster. Number of mergers per Gyr and per single cluster as a function of the cluster velocity dispersion. Symbols and colors are the same as in panel **a**. Dotted lines represent best-fit power-law functions. The ratio between the Chi-square and the number of degrees of freedom is $\chi^2/NDF < 2.6$ for NSSTBH and $< 0.5$ for BHSTNS, assuming a 10% error associated to the measure of the scattering rate and the measure of the merging probability.

Under these assumptions we find a NS-BH merger rate

$$\Gamma_{GC} = (3.2 - 8.5) \times 10^{-3} \times (\rho_{MWEG} / 0.0116\ \mathrm{Mpc^{-3}})\ (N_c/200)\ \mathrm{yr^{-1}\ Gpc^{-3}} \quad \text{for NSSTBH}, \qquad (3)$$
$$\quad\quad (1.7 - 2.5) \times 10^{-1} \times (\rho_{MWEG} / 0.0116\ \mathrm{Mpc^{-3}})\ (N_c/200)\ \mathrm{yr^{-1}\ Gpc^{-3}} \quad \text{for BHSTNS}, \qquad (4)$$

with the two extremes corresponding to different metallicities. We find a remarkably well agreement with very recent results based on a sample of ~140 Monte Carlo simulations of GCs[37]. Regarding galactic nuclei, the mass and half-mass radius of the Galactic NC[38] are $M_{GC} = 2.2 \times 10^7\ M_\odot$ and $R_{GC} \sim 5$ pc, respectively, thus corresponding to $\sigma = 40$-$60$ km/s.

Under these assumptions, the merger rate for NCs in Milky Way analogs is

$$\Gamma_{NC} = (0.9 - 1.7) \times 10^{-4} \times (\rho_{MWEG} / 0.0116 \text{ Mpc}^{-3}) \text{ yr}^{-1} \text{ Gpc}^{-3} \text{ for NSSTBH}, \quad (5)$$
$$(1.0 - 1.5) \times 10^{-2} \times (\rho_{MWEG} / 0.0116 \text{ Mpc}^{-3}) \text{ yr}^{-1} \text{ Gpc}^{-3} \text{ for BHSTNS}. \quad (6)$$

Our estimates rely upon the assumption that all Milky Way-like galaxies harbor an NC, thus they represent an upper limit to the actual merger rate.

Note that the merger rates for NCs and GCs are comparable for configuration BHSTNS, thus suggesting that NCs might account for 10-20% of the total population of dynamical NS-BH mergers. Figure 4 shows the variation of $\Gamma_c$ as a function of the cluster mass and half-mass radius. Our results are superimposed to the sample of observed GCs[29] and NCs[30]. Only the heaviest and more compact NCs can sustain at least 1 event per Gyr. Table 1 summarizes all the models investigated, highlighting the number of exchanges -- an exchange marks the formation of a NS-BH -- , the number of mergers, and the number of possible EM counterpart out of 10000 simulations.

We stress that none of the observed clusters in the sample exhibit any evidence of a central massive BH (MBH), neither supermassive -- for NCs -- nor of intermediate-mass -- for GCs.

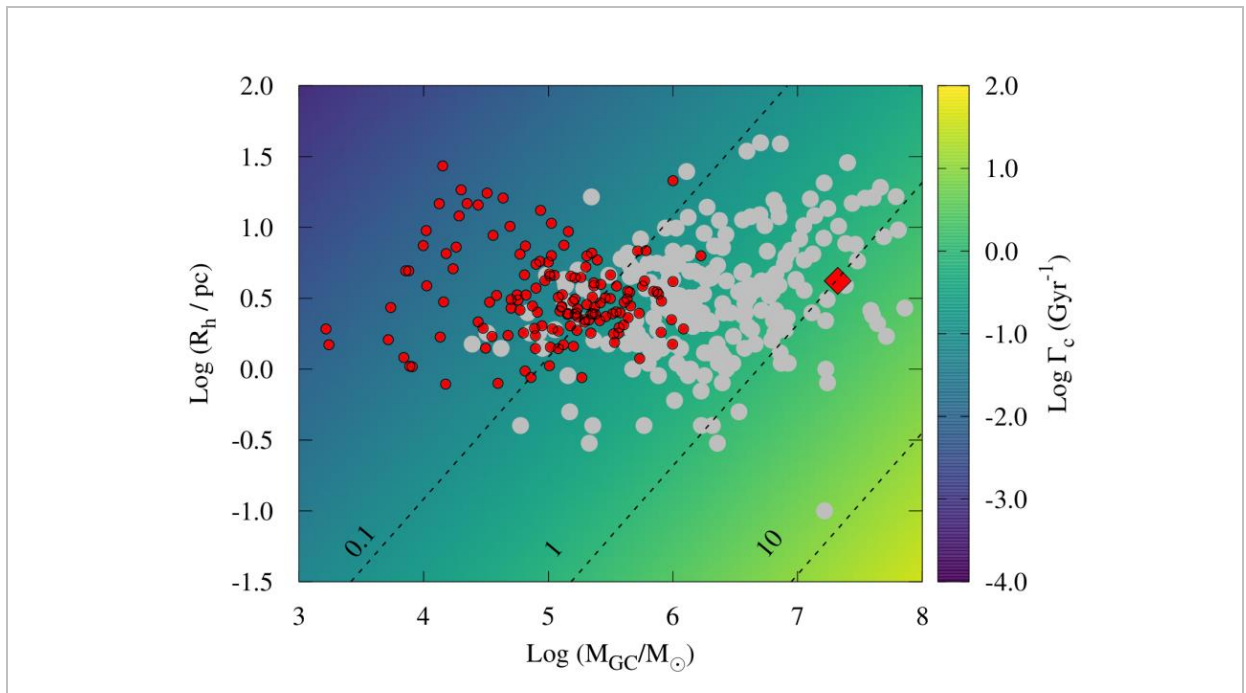

Fig 4. Merger rate per cluster. The surface map shows how the merger rate varies at varying the logarithm cluster half-mass radius ($R_h$) in parsecs and cluster mass ($M_{GC}$) in solar masses. Colours define the amount of mergers per year ($\Gamma_c$). *Red circles represent Galactic globular clusters[29], while grey points represent a sample of local nuclear clusters[30] and the red diamond identifies the Galactic nuclear cluster[38]*. Dotted lines mark the limit of 0.1, 1, and 10 events per Gyr per cluster, respectively.

Inside the so-called influence radius, $R_{inf}$, it is possible to show that the relaxation time for MBH with masses in the range $10^4$ - $10^9$ $M_\odot$ is similar to $t_{rel}$ of clusters with masses in the same mass range, thus NS-BH formation could proceed similarly to NCs. However, deep into the influence radius, where the mass budget is dominated by the MBH itself and the velocity dispersion scales with $R_{inf}^{-1/2}$, the relaxation time will increase as $R_{inf}^{3\delta/2}$, being $\delta > 0$ a factor that depends on the matter distribution around the MBH. For $\delta = 1$, this implies that the relaxation time inside $0.1 R_{inf}$ exceeds a Hubble time if $M_{MBH} > 10^6$ M$_\odot$, thus indicating that the NS-BH formation channel explored here could be strongly suppressed in heavy galactic nuclei. The late evolution of a NS-BH binary formed around an MBH will depend on a number of processes. First, due to mass segregation, the binary will migrate inward, passing through regions at increasing density and velocity dispersion. This corresponds to a reduction of the hard-binary separation, meaning a larger probability for the binary to be disrupted if its hardening rate is not sufficiently large. Second, the increasing gravitational torque associated with the MBH can tidally rip apart the binary. Third, if the binary survive to both energetic scatterings and tidal torques, the reduced distance to the MBH could onset Kozai-Lidov oscillations[39,40], which can excite the binary eccentricity up to unity potentially shortening its lifetime. Quantifying these effects for SMBHs is challenging owing to the fact that the physics regulating star formation and dynamics around an SMBH is still not fully understood. For IMBHs in star clusters, this is even more difficult, owing to the lack of conclusive evidence of their existence and of a well-constrained formation scenario. For instance, recent numerical models suggest that IMBHs forming out of a sequence of stellar collisions are associated with clusters retaining only one or two BHs after the IMBH growth, thus limiting the probability for the NS-BH dynamical channel presented here to take place.

Besides the formation of NS-BH mergers, we find in the case of configuration BHSTNS that the NS flyby can push the stellar companion on an orbit passing sufficiently close to the BH to trigger the stellar disruption and associated tidal disruption event (TDE). The probability for this to happen increases at increasing the velocity dispersion, being ~1% for metal-poor and 1.5% for metal-rich clusters with $\sigma = 5$ km/s. The scattering rate for these events is larger for metal-poor systems, as here the BH mass is larger, resulting in a larger cross section and, thus, in a larger scattering rate. For values typical of Milky Way GCs, the resulting TDE rate is $\Gamma_{TDE}$ ~ (2.4 - 4.2) x $10^{-9}$ yr$^{-1}$, compatible with the value expected for TDEs triggered by BH binaries[41,42]. The 90% of disrupted stars have a mass < 0.5 M$_\odot$, thus possibly representing white dwarfs or low-mass main sequence stars.

| Model | Z (Z$_\odot$) | $\sigma$ (km/s) | | | | | |
|---|---|---|---|---|---|---|---|
| | | 5 | 15 | 20 | 35 | 50 | 100 |
| | | Number of **exchanges** - **mergers** - Electromagnetic counterparts in 10000 simulations | | | | | |
| NSSTBH | $10^{-2}$ | 641  11  2 | 769  16  4 | 795  22  3 | 861  30  9 | 911  37  14 | 950  83  34 |
| | 1 | 725  9  0 | 880  19  2 | 971  24  7 | 1001  38  12 | 1089  64  15 | 1021  103  19 |
| BHSTNS | $10^{-2}$ | 200  3  3 | 215  6  3 | 268  13  9 | 274  22  17 | 311  30  21 | 326  77  51 |
| | 1 | 280  4  1 | 341  13  6 | 365  21  10 | 430  37  19 | 460  56  25 | 462  116  45 |

**Tab 1.** Number of NS-BH mergers formed in our models. We refer to two configurations, either a neutron star – star binary impacting over a black hole (configuration NSSTBH) or viceversa (BHSTNS), metallicity Z (in unit of solar values), and velocity dispersion $\sigma$ (in unit of kilometers per

second). Each set consists of 10000 simulations.

**Identifying dynamical NS-BH mergers with GW emission.** According to the forefront of binary stellar evolution recipes[18], BH in isolated NS-BH mergers are expected to feature masses strongly peaked in the range 6.5 - 8.5 $M_\odot$ and NS masses broadly distributed between 1.4 - 2 $M_\odot$, thus corresponding to chirp masses < 4 $M_\odot$. Figure 5a shows the chirp mass, $\mathcal{M}_{chirp}$, distribution for all our high-velocity dispersion dynamical models. We refer to models with $\sigma$ = 100 km/s to discuss the general properties of dynamical mergers. This choice is motivated by the larger number of mergers for these models, which allow a more robust statistical investigation of merger mass distribution. Nonetheless, the overall distribution shown in the following does not differ from those at smaller $\sigma$ values, although the latter are affected by a lower statistics. Mergers forming dynamically in our simulations, instead, show a non-negligible probability to have larger $\mathcal{M}_{chirp}$. In configuration NSSTBH, up to 52%(32%) of mergers in metal-poor(rich) clusters have a chirp mass above this threshold. The percentage decreases for BHSTNS configuration but is still not negligible, being 14 - 17%, with the lower limit corresponding to metal-rich systems. A chirp mass above 4 $M_\odot$ thus represents the first clear distinctive mark of an NS-BH merger with a dynamical origin.

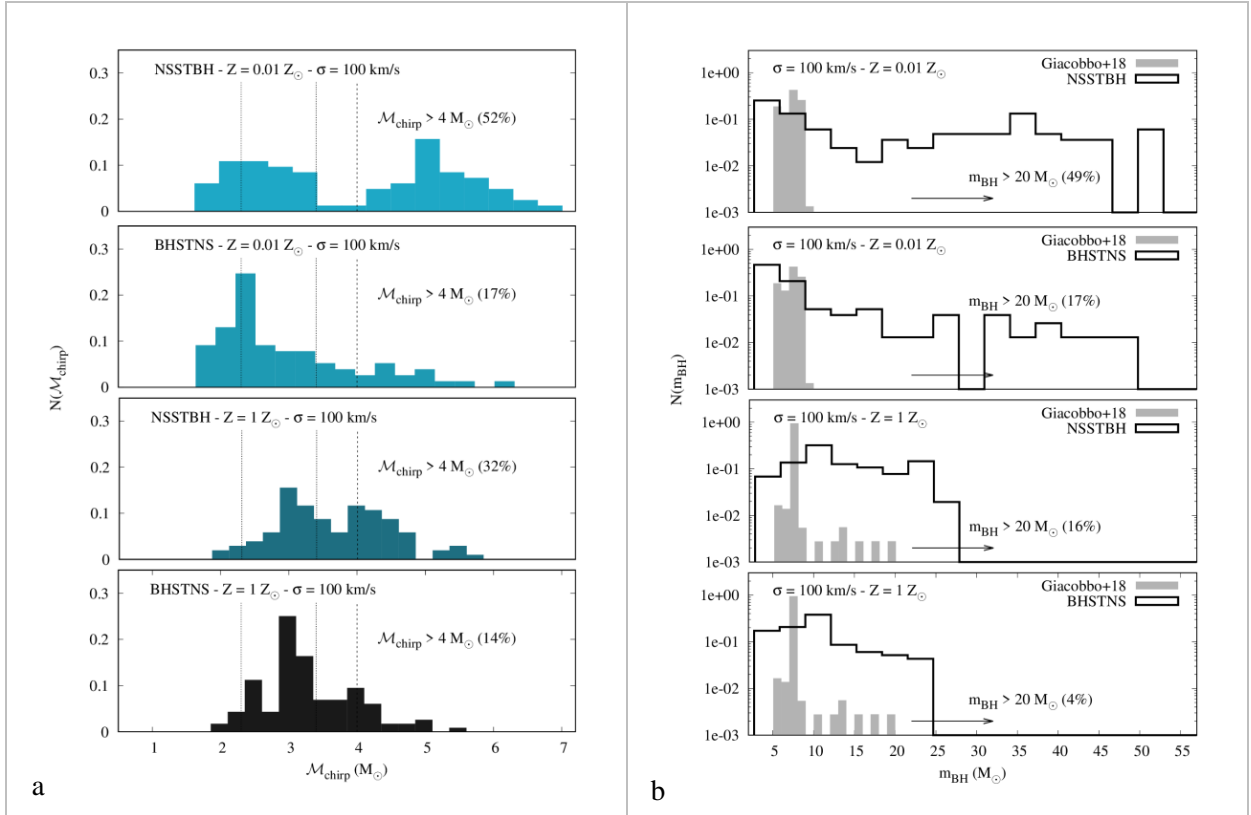

Fig 5. Comparison between chirp and BH mass of isolated and dynamical mergers **a.** Chirp mass distribution for dynamical NS-BH mergers. Different colors correspond to different configurations (either a binary neutron star – star and a single black hole, NSSTBH, or viceversa, BHSTNS), velocity dispersion ($\sigma$) and metallicity ($Z$). The vertical dotted lines set the range allowed for isolated mergers. We provide the percentage of models with a chirp mass $\mathcal{M}_{chirp} > 4$ $M_\odot$, indicated by the dashed line. **b.** Mass distribution of dynamical NS-BH mergers. The distribution of BH masses ($m_{BH}$) is shown for different configurations, metallicity, and velocity dispersion. Open steps identify dynamical mergers, whereas filled grey step represent isolated mergers as calculated in Giacobbo et al (2018)[18]. Labels indicate the probability to obtain a NS-BH merger with black hole mass $m_{BH} > 20$ M$_\odot$ for different configurations.

By definition, a large chirp mass indicates a large binary mass and, in the case of NS-BH binaries, this can indicate a large BH mass. In fact, the second characteristic mark of dynamical mergers is apparent in the BH mass distribution shown in Figure 5b. For clarity's sake, we overlay to our predictions the same quantity inferred for isolated NS-BH mergers[18]. In metal-poor clusters, we find that more than 50% of NSSTBH and 17% of BHSTNS simulations lead to a merger involving a BH with mass $m_{BH} > 20$ M$_\odot$. The percentage drops to 16% and 4%, respectively, for metal-rich clusters, due to the lower maximum BH mass set by stellar evolution for metal abundances close to solar (see the Method section for further details about the initial BH mass spectrum). However, we note that in comparison to isolated mergers, which predicts a narrow peak at $m_{BH} \sim 7$ M$_\odot$, dynamical mergers show a broad distribution even in the mass range 10-20 M$_\odot$, thus suggesting that the dynamical channel could dominate over isolated binaries already in this BH mass range. A BH mass above 10 M$_\odot$ thus represents the second distinctive mark of a dynamical origin for NS-BH mergers.

**Electromagnetic counterparts.** One of the most interesting outcomes of a NS-BH merger is the possible development of an EM counterpart. This is associated with an accretion disc formed from NS debris during the merging phases. The disc can form only if the BH tidal field torns apart the NS before it enters the BH event horizon, a condition fulfilled if the NS tidal radius

$R_{tid} = R_{NS} (3m_{BH}/m_{NS})^{1/3}$, (7)

exceeds the BH's ISCO[39]

$R_{ISCO} = Gm_{BH}/c^2 [3 + Z_2 - \text{sign}(\chi)(3 - Z_1)(3 + Z_1 + 2Z_2)]^{1/2}$, (8)

where $Z_{1,2}$ are functions of the BH adimensionless spin parameter $\chi = a_{BH}/m_{BH}$. Therefore, the merger will not feature associated EM counterpart if $R_{tid}/R_{ISCO} < 1$. Note that the opposite does not represent a conditio sine-qua-non for the development and detectability of an EM signal, as in the case $R_{tid}/R_{ISCO} > 1$ this depends on the geometry of the merger with respect to the observer and other potential observational biases. Figure 6 shows how the $R_{tid}/R_{ISCO}$ ratio varies at varying $m_{BH}$, assuming a NS radius[3] $R_{NS} = 12$ km and mass $m_{NS} = 1$-$3$ M$_\odot$, and different $\chi$ values. For mildly rotating BHs ($\chi \sim 0.5$), mergers meet the condition to enable EM emission only if the BH has a mass $m_{BH} < 3.8$ M$_\odot$. In this case, neither the isolated channel nor the dynamical are expected to be prone to EM emission, being the mass of merging BHs larger than this threshold. For spin values similar to those inferred from LIGO observations[1] ($\chi \sim 0.7$), the threshold BH mass shifts to 5.2 M$_\odot$. In this case, we find 15 mergers out of 854 merger candidates, regardless of the configuration, with mass below this threshold, thus the probability to develop an EM counterpart is limited to $< 1.8\%$. For highly spinning BHs ($\chi \sim 0.9$), instead, the BH mass threshold is 9.2 M$_\odot$. In this case, the vast majority of mergers in the isolated channel, especially for metal-poor environments, will fall in the region where EM counterpart is allowed, whereas dynamical mergers have a probability of 53.4% to fall outside this threshold, thus implying the impossibility for an EM counterpart to develop. Thus, the absence of a clear EM counterpart with a high-spin BH represents the third clear mark of a dynamical origin.

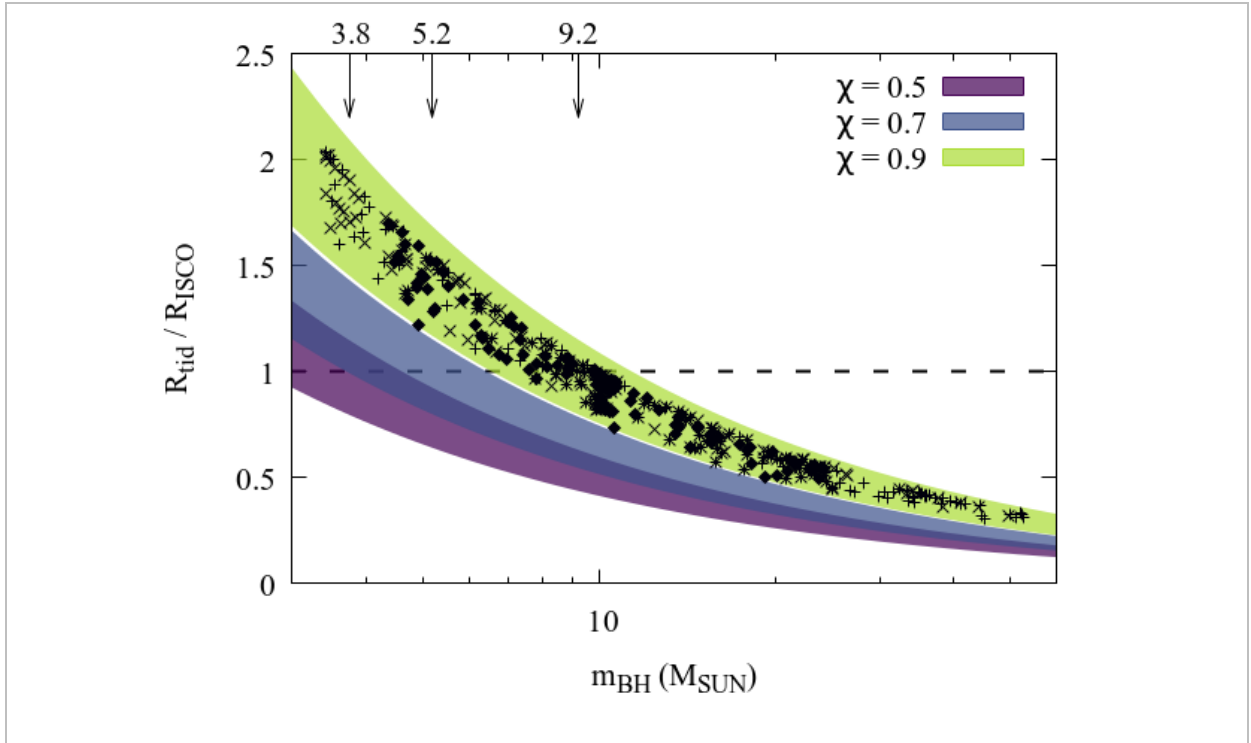

Fig 6. Limits on electromagnetic counterpart to a neutron star - black hole merger. Ratio between the neutron star tidal radius ($R_{tid}$) and the black hole innermost stable circular orbit (ISCO, $R_{ISCO}$) and the black hole mass ($m_{BH}$, x-axis) for mergers in all models with a velocity dispersion $\sigma = 100$ km s$^{-1}$. Coloured regions enclose the limiting values of $R_{tid}/R_{ISCO}$ assuming a spin parameter $\chi = 0.5, 0.7, 0.9$. Arrows and accompanying labels mark value of $m_{BH}$ above which an electromagnetic (EM) counterpart cannot develop, with smaller values corresponding to lower spins. Points represent all mergers in our models, with different symbols representing different sets. Only points lying above $R_{tid}/R_{ISCO} < 1$ can give rise to an EM counterpart.

**Eccentricity distribution and prospect for multiband GW observations**

Looking at the eccentricity distribution prior to the scattering and after, and restricting the analysis to the cases that eventually lead to a merger, we find that at formation dynamical mergers are characterized by an extremely narrow eccentricity distribution peaked around unity. To explore whether some residual eccentricity is preserved when the merger enters the frequency bands of interest for GW detection we calculate the evolution of the GW characteristic strain as a function of the frequency for all mergers, assuming that they are located in the local Universe, at a luminosity distance $D_L = 230$ Mpc (redshift = 0.05). Note that this is compatible with the luminosity distance inferred for the two NS-BH merger candidates reported in the GRACE-DB. Fig. 7a shows the eccentricity distribution as binaries cross the $10^{-3}, 10^{-2}, 10^{-1}, 10^{0}, 10^{1}$ Hz frequency bands. Note that a large fraction of binaries have $e > 0.1$ in mHz, i.e. in the observation band of space-based detectors like LISA, but none of them have $e > 0.1$ when crossing the 1 Hz frequency threshold. Nonetheless, dynamical mergers appear to be potential multiband GW sources in the 0.01-1 Hz frequency range. Fig. 7b shows the characteristic strain of mergers with a total merger time shorter than $10^5$ yr in all our models. We overlap to the simulated sources the sensitivity curve -- in terms of characteristic strain -- for low-frequency GW detectors (LISA, DOs[5], ALIA[40,41], and DECIGO[42]) and high-frequency detectors (LIGO, KAGRA, and the Einstein Telescope). Decihertz observatories would constitute precious instruments to follow the evolution of these sources during the in-spiral phase down to the merger. In the same plot, we show an example for the signal of a merger taking place within the Andromeda galaxy, located at a distance of ~ 779 kpc, or the Virgo galaxy cluster (~ 20 Mpc). Mergers occurring at distances between Andromeda and the Virgo cluster could spend enough time in the LISA band to be detected several years prior to the merger.

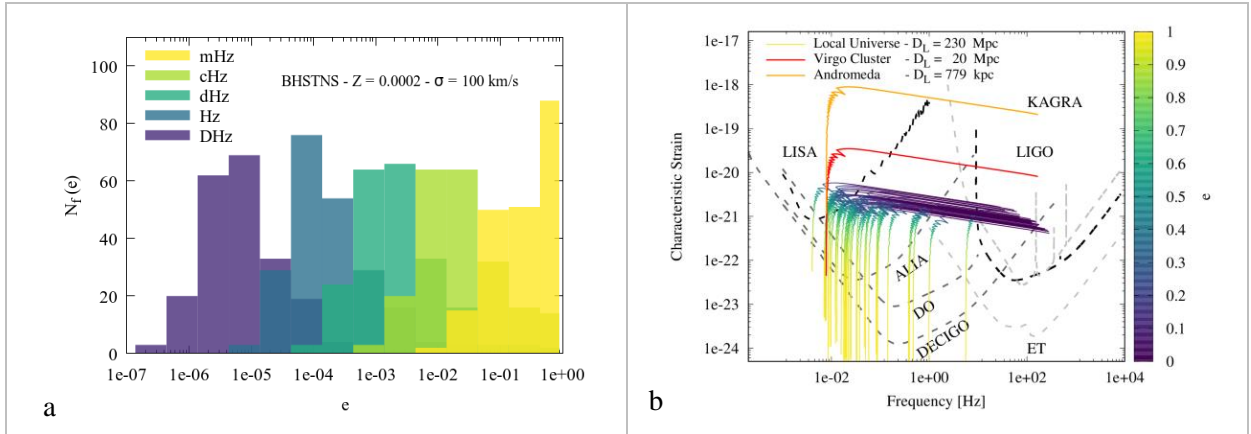

Fig 7. Mergers properties during the inspiral. **a.** Mergers eccentricity distribution. The plot shows the eccentricity distribution for all mergers for the configuration with a single neutron star and a black hole – star binary (BHSTNS) assuming a metallicity value typical of old star clusters ($Z = 0.0002$) and a velocity dispersion $\sigma = 100$ km s$^{-1}$. Different colors correspond to different frequency bands, namely $10^{-3}$ (millihertz, mHz), $10^{-2}$ (centihertz, cHz), $10^{-1}$ (decihertz, dHz), $10^0$ (hertz, Hz), $10^1$ (decahertz, DHz). The eccentricity is measured as the merger crosses such bands. **b.** Mergers gravitational waves strain. Characteristic strain as a function of frequency for all mergers with a total merger time smaller than $10^5$ years. Colour-coding marks the eccentricity ($e$) variation. All coloured tracks are assumed to be located at a luminosity distance of $D_L = 230$ Mpc (redshift = 0.05). The characteristic strain for one single example is also shown assuming the source is located in the Virgo Cluster (red straight line) or in the Andromeda Galaxy (orange straight line).

**Discussion**

We modelled the dynamical formation of NS-BH mergers in massive clusters, exploring the phase space in terms of cluster velocity dispersion and metallicity, and assuming different configurations. We infer an optimistic merger rate of $\Gamma_{GC} = 0.1$ yr$^{-1}$ Gpc$^{-3}$ for GCs and $\Gamma_{NC} = 0.01$ yr$^{-1}$ Gpc$^{-3}$ for NCs, much lower than the rate inferred after the first two LIGO observational campaigns[1] ($< 610$ yr$^{-1}$ Gpc$^{-3}$). This might indicate that dynamical mergers bring a little contribute to the overall population of NS-BH mergers. Nonetheless, our models suggest that dynamical mergers can exhibit distinctive marks potentially useful to interpret GW observations. While the isolated channel predict mergers with BH masses strongly peaked around 7 M$_\odot$, and chirp masses below 4 M$_\odot$, a non-negligible percentage of dynamical mergers could be characterised by BH masses above 20 M$_\odot$ and chirp masses above 4 M$_\odot$. This difference has important implications for the development of an EM counterpart. For highly spinning BHs (spin $\chi = 0.9$), the isolated channel suggests that all mergers have the possibility to produce coincident EM+GW emission. Conversely, in the dynamical channel up to 50% mergers have BH masses $> 10$ M$_{SUN}$, sufficiently large to avoid the NS disruption outside the BH ISCO. We conclude that a dynamical merger might be uniquely identified if it fulfills simultaneously the requirements that: i) the chirp mass exceeds 4 M$_\odot$, ii) the BH mass exceeds 20 M$_\odot$, and

iii) an EM is absent if the BH spin exceeds 0.9. Dynamical NS-BH mergers appear to be promising multiband sources that might be observable with future decihertz detectors. Exceptional cases could be observed even with LISA, provided that the merger occurred at distances ~ 0.7 - 20 Mpc, like in Andromeda or in the Virgo Galaxy cluster.

**Methods**

**Comparing observations and numerical models.** To compare our models with observations, we exploit the catalog of Milky Way GC[29], which provides, among other quantities, the distribution of velocity dispersion ($\sigma$), half-mass radius ($R_{GC}$), and relaxation time ($t_{rel}$), as shown in Figure 8. Typical values for Galactic GCs are, in general, $\sigma$ ~ 4-6 km/s, $R_{GC}$ ~ 1-5 pc and $t_{rel}$ ~ 1 Gyr.

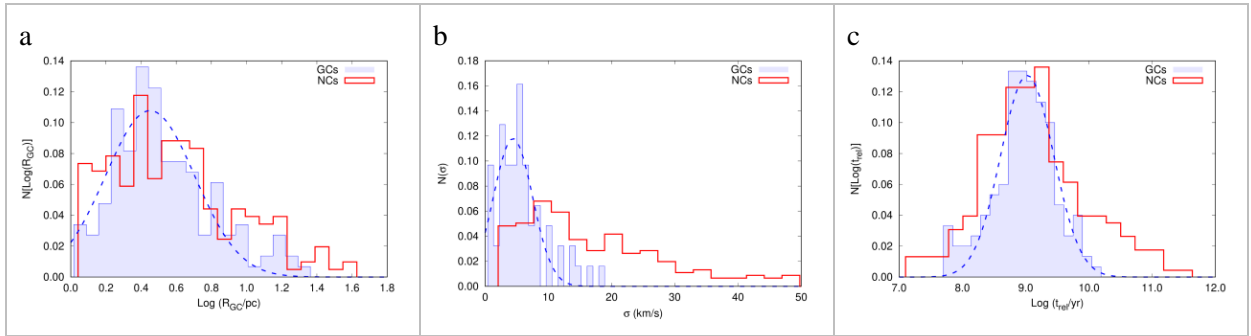

Fig 8. Properties of Galactic GCs and nearby NCs. **a.** Clusters half-mass radius. Distribution of half-mass radius ($R_{GC}$) in parsecs for Milky Way globular clusters (GC, shaded steps) and nuclear clusters observed in the local Universe (NC, red steps). The dotted line represents the best-fit Gaussian distribution. **b.** Clusters velocity dispersion. Distribution of clusters velocity dispersion ($\sigma$) in kilometer per second for the same sample of clusters shown in panel a. **c.** Clusters relaxation time. Distribution of clusters relaxation time in yr for the same sample of clusters shown in panel a.

As shown in Figure 9, the GC mass, half-mass radius, and velocity dispersion are connected by a tight relation in the form

$$\text{Log } GM_{GC}/R_{GC} = \alpha_b + 2\text{Log } \sigma \,. \tag{9}$$

Using a least square fit we find $\alpha_b = 1.14 \pm 0.03$, with an associated ratio between the $\chi^2$ and the number of degrees of freedom $\chi^2/\text{NDF} = 0.062$.

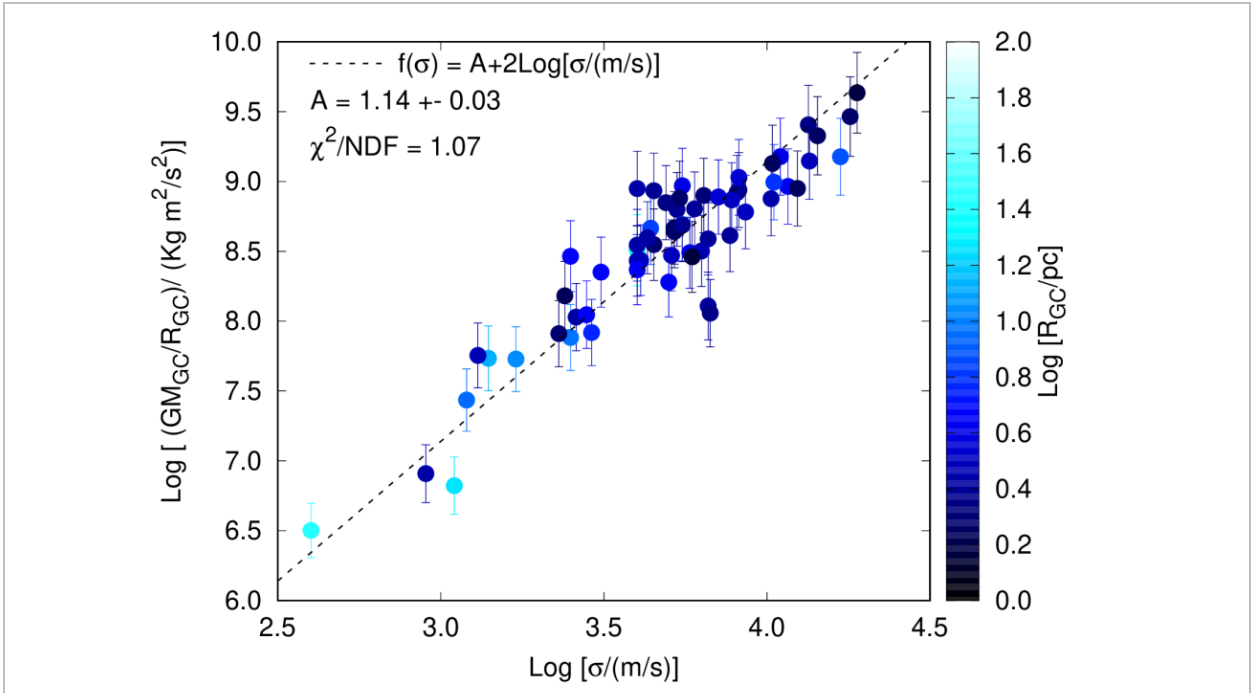

Fig 9. Cluster mass-radius-velocity dispersion relation. Scaling relation connecting the cluster mass ($M_{GC}$) and half-mass radius ($R_{GC}$) to the cluster velocity dispersion ($\sigma$). Note that the quantity shown on the y-axis represents a proxy to the cluster potential energy ($GM_{GC}/R_{GC}$). Colours identify the clusters half-mass radius. Data represent Galactic globular clusters for which all the three quantities are available (as detailed in the Harris (2010) catalog[29]). The best-fit is obtained assuming a 0.1dex error on the y-axis.

We use Equation 9 to convert the velocity dispersion, which is an input parameter in our N-body simulations, into GC mass and half-mass radius. We use the same strategy to compare our results with a sample of 228 NCs observed in the local Universe[30], exploiting published mass and half-mass radius to calculate the velocity dispersion and half-mass relaxation time (see Figure 8).

**Setup of the *N*-body simulations and numerical approach.** The direct *N*-body simulations presented in this work have been performed exploiting the ARCHAIN code[27,43,44], which features a treatment for close encounters called algorithmic regularization[43] and includes General Relativity effects via post-Newtonian formalism[27] up to order 2.5. The choice of modelling a compact object paired with a stellar companion is twofold. On the one hand, stars constitute 90% of the total stellar population in a star cluster, making probable for them to be captured by a heavier object. On the other hand, since stars are lighter than compact objects it is energetically convenient for a binary to exchange components and increase its binding energy. Heavy compact binaries in star clusters can indeed form via a sequence of such interactions[45,46], which indeed can contribute to the formation of NS-BH in star clusters[19]. To initialize the BH and NS masses, we sample the zero-age main sequence mass of the three components assuming a power-law mass function[47], namely $f(m_*) \propto m_*^{-2.3}$. We calculate the remnant masses taking advantage of the SSE tool[48] for NSs and state-of-the-art mass spectra[26] for BHs. The latter are used because stellar evolution recipes for massive stars implemented in SSE are outdated. We show the BH mass spectrum adopted in Figure 10a. Note that at low metallicities, the mass of the BHs extends to up to 60 $M_\odot$, while being much smaller for metal-rich progenitors. This is at the basis of the differences between the results obtained for different configurations with different metallicity values.

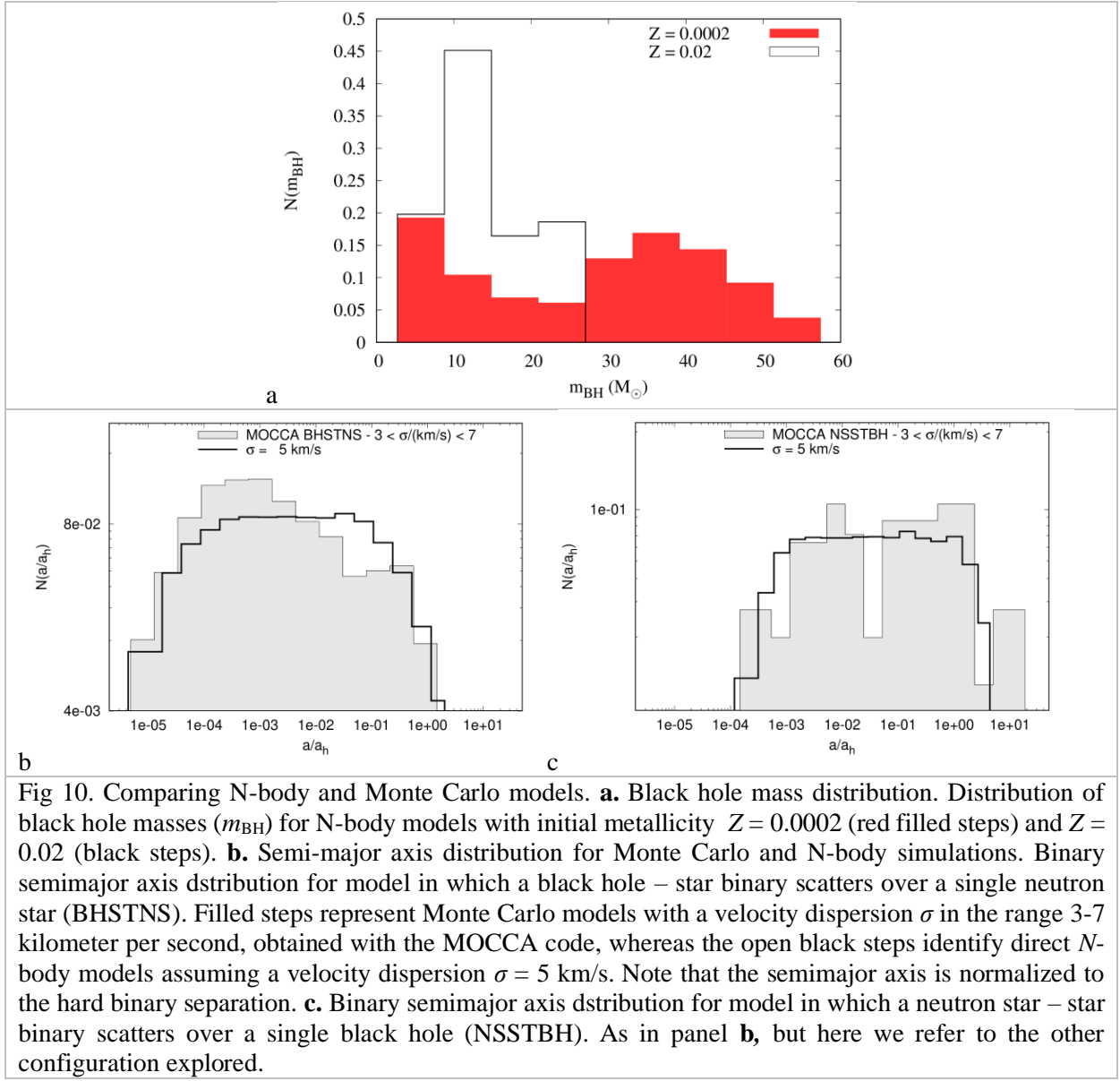

Fig 10. Comparing N-body and Monte Carlo models. **a.** Black hole mass distribution. Distribution of black hole masses ($m_{BH}$) for N-body models with initial metallicity $Z = 0.0002$ (red filled steps) and $Z = 0.02$ (black steps). **b.** Semi-major axis distribution for Monte Carlo and N-body simulations. Binary semimajor axis dstribution for model in which a black hole – star binary scatters over a single neutron star (BHSTNS). Filled steps represent Monte Carlo models with a velocity dispersion $\sigma$ in the range 3-7 kilometer per second, obtained with the MOCCA code, whereas the open black steps identify direct *N*-body models assuming a velocity dispersion $\sigma = 5$ km/s. Note that the semimajor axis is normalized to the hard binary separation. **c.** Binary semimajor axis dstribution for model in which a neutron star – star binary scatters over a single black hole (NSSTBH). As in panel **b,** but here we refer to the other configuration explored.

We note that a smoother mass function would lead to a larger population of massive BHs. This can lead to an increase of the probability for BHs in NS-BH mergers to have a mass larger than the value typical for isolated binaries (~ 7 M☉). This, in turn, would increase the amount of dynamical mergers that might be clearly distinguishable from the isolated ones.

We assume that the three-body interaction is hyperbolic and in the regime of strong deflection, namely the outer angle between the incoming and outcoming direction of the scattering object is larger than 90 degrees, and the maximum pericentral distance between the binary and the third object equals twice the binary semi-major axis. We restrict our analysis to strong scatterings as these are the only capable to trigger an exchange between one of the binary components and the third object. To set the maximum semimajor axis $a$ we follow recent numerical results showing that this quantity divided by the binary reduced mass μ is proportional to the ratio between the host cluster half-mass radius and total mass[49], namely $a/\mu = kR_{GC} / M_{GC}$. The scaling constant $k = 1/54$ claimed in literature is typical of dynamically processed binaries, i.e. that underwent several dynamical encounters, while here we focus on binaries not fully dynamically processed. To mimic this assumption we set $k = 10$, and we calculate the $R_{GC} / M_{GC}$ ratio through $\sigma$ via Equation 9. If $a$ calculated this way is larger than the distance below which the star gets torn apart by tidal torques exerted by the companion, we set as maximum value allowed the hard binary separation $a_h$. The minimum binary separation is set as the maximum between 100 times the star's Roche lobe or 1000 times the ISCO of the compact object in the binary. This avoids the possibility that the star plunges inside the BH or is immediately disrupted before the scattering takes place. We initialize our scattering experiments basing our assumptions on previously published works focused on Monte Carlo modelling of GCs. To check the consistency of our assumptions, we compare the distribution of the binary semimajor axis normalized to the hard binary separation in our models with $\sigma = 5$ km/s and MOCCA simulations, as shown in panels b and c of Figure 10 . This quantity seems well suited to compare the two approaches, as it contains information about binaries orbital properties, via the semimajor axis and the component masses, and their hosting environment via $\sigma$. We find that, in general, the adopted distribution does not deviate dramatically from MOCCA results, thus providing an acceptable compromise that allows us to expand the study beyond the capability of MOCCA models.

The initial binary eccentricity is sampled from a thermal distribution. Initial velocities of the binary and the single object are taken assuming a Maxwellian distribution of the velocities characterized by the star cluster velocity dispersion $\sigma$. We assume $\sigma = 5, 15, 20, 35, 50, 100$ km/s and two values for the stellar metallicity, either $Z = 2 \times 10^{-4}$, typical of old GCs, or solar values ($Z = 0.02$). As summarized in Table 1, our models can be divided into two main set, depending on the scattering configuration, each set is divided into two subsets, depending on the metallicity, and each subset is divided into 6 simulations sample depending on $\sigma$. Thus, we gather a total of 24 simulation samples each consisting of 10000 simulations.

**Calculating the scattering rate for *N*-body models.** The interaction rate can be written as $dR/dt = n\sigma\Sigma$, where $n$ is the density of scattering particles, $\sigma$ is the velocity dispersion, and $\Sigma$ is the binary cross section

$$\Sigma = \pi a^2 (1-e)^2 [1 - 2G(m_1+m_2+m_3)/(\sigma^2 a (1-e))] . \qquad (10)$$

For each simulation set, we calculate $\Sigma$ by using the median value of $a$, $e$, $m_1$, $m_2$, and $m_3$, calculated from the assumed initial distribution. The number density $n$ of scattering particles depends critically on the amount of NSs and BHs left in the cluster. For BHs, we exploit our recent studies on BH subsystems in GCs[22,23]. Using MOCCA models, we find that the typical density of the BH ensemble is comparable to the cluster density, $n_{BH} \equiv n \approx M_{GC}/(m_* R_{GC}^3)$. For NSs, instead, we consider the fact that segregation is mostly prevented by the BHs present in the cluster, whereas NS-to-total mass ratio for typical clusters is of the order of 0.01, a limit imposed by the standard initial mass function. Thus, we assume $n_{NS} \equiv 0.01n$ as an upper limit to the actual NS number density. Under these assumptions, we derive an optimistic estimate of the scattering rate $dR/dt$ that, for $\sigma = 5$ km/s, results into:

$$dR/dt \ (5 \text{ km/s}) = \ 2 - 4 \quad \text{Gyr}^{-1} \quad \text{for NSSTBH,} \qquad (11)$$
$$= 150 - 400 \text{ Gyr}^{-1} \quad \text{for BHSTNS.} \qquad (12)$$

Note that these estimates fall in the range of values derived from self-consistent MOCCA models, for which we find 6.3 Gyr$^{-1}$ and 254 Gyr$^{-1}$, respectively.

**Data Availability.** The data sets generated during the current study are available from the corresponding author on reasonable request. The updated version of the ARCHAIN code used to carry out the *N*-body simulations is available from the corresponding author on reasonable request.

**Acknowledgements**
The author is grateful to Mirek Giersz, Abbas Askar, and Arkadiusz Hypki for providing access to the MOCCA database and for their invaluable help in managing the data. The author acknowledges financial support from the Alexander von Humboldt Foundation and the Federal Ministry for Education and Research in the framework of the research project "The evolution of black holes from stellar to galactic scales". The author acknowledges support from the Deutsche Forschungsgemeinschaft (DFG, German Research Foundation) -- Project-ID 138713538 -- SFB 881 ("The Milky Way System"), and the COST Action CA16104 "GWVerse". The author acknowledges the Baden-Wurttemberg HPC infrastructure for providing access to the bwForCluster supercomputer, and to the Astronomisches Rechen Institut for providing access to the Kepler cluster, where simulations were carried out.
**Competing Interests.** The author declare no competing interests.
**Contributions** The author had the idea of studying the channel investigated here, he carried out the analysis on the MOCCA database and ran the *N*-body simulations, performing the analysis of the results. The author wrote the text.
**Correspondence.** Correspondence and requests for materials should be addressed to Manuel Arca Sedda (email: m.arcasedda@gmail.com).

**Supplementary material**